# Robust Transceiver Design for Reciprocal $M \times N$ Interference Channel Based on Statistical Linearization Approximation

Ali D. Mayvan [a], Hassan Aghaeinia [b], Mohammad Kazemi [c]

[a b c] *Department of Electrical Engineering, Amirkabir University of Technology (Tehran Polytechnic), Tehran, Iran.*

[a] *Email: alidalir@aut.ac.ir*

[b] *Email: aghaeini@aut.ac.ir*

[c] *Email: mohammadkaazemi@aut.ac.ir*

Corresponding Author: alidalir@aut.ac.ir

## *Abstract*

This paper focuses on robust transceiver design for throughput enhancement on the interference channel (IC), under imperfect channel state information (CSI). In this paper, two algorithms are proposed to improve the throughput of the multi-input multi-output (MIMO) IC. Each transmitter and receiver has respectively **M** and **N** antennas and IC operates in a time division duplex mode. In the first proposed algorithm, each transceiver adjusts its filter to maximize the expected value of signal-to-interference-plus-noise ratio (SINR). On the other hand, the second algorithm tries to minimize the variances of the SINRs to hedge against the variability due to CSI error. Taylor expansion is exploited to approximate the effect of CSI imperfection on mean and variance. The proposed robust algorithms utilize the reciprocity of wireless networks to optimize the estimated statistical properties in two different working modes. Monte Carlo simulations are employed to investigate sum rate performance of the proposed algorithms and the advantage of incorporating variation minimization into the transceiver design.

## 1. Introduction

Normally, wireless network scenarios such as interference channel (IC), share the channel among the users, resulting in multi-user interference. A new method, termed interference alignment (IA), leads to the efficient use of communication resources, since it successfully achieves the theoretical bound on the multiplexing gain [1-3]. In this scheme, unwanted signals from other users are fitted into a small part of the signal space observed by each receiver, called interference subspace. The other signal subspace is left free of interference for the desirable signal. The performance of IA scheme is sensitive to channel state information (CSI) inaccuracies. In [4], and [5] the performance of IA under CSI error was quantified. Asymptotic mean loss in sum rate compared to the perfect CSI case was derived [4]. Multiplexing gain is fully achievable when the variance of the CSI measurement error is inversely proportional to the SNR [4]. Performance of IA for SISO and multi-input multi-output (MIMO) IC where the transmitters are provided with the quantized CSI via feedback can be found in [6] and [7]. Other feedback strategies have been presented in [8]. The IA problem has been studied for the cognitive radio networks in [9-11]. The performance of ad-hoc networks using the IA has been evaluated in [12].

*1.1. Motivation*

In order to maximize sum rate of the MIMO interference network, a beamforming strategy based on the interference alignment is used. Such algorithms have been established by progressive minimization of the leakage interference [13, Algorithm 1] and [14]. The Max-SINR algorithm [13, Algorithm 2], minimum mean square error [15] and joint signal and interference alignment [16] are other algorithms. These schemes are established based on availability of perfect CSI. The performance of transceivers is degraded by the CSI error. Different algorithms are proposed to improve the throughput of the IC, under imperfect CSI. In the literature, precoder-decorrelator optimization is proposed for broadcast and point-to-point systems [17–20]. Specifically, in [17], and [20] the authors consider precoding design for multi-input single-output (MISO) broadcast channel, where it is shown that the precoder optimization problem is always convex. In [19] the authors consider precoder-decorrelator design for MIMO broadcast channel using an iterative algorithm based on solving convex sub-problems. On the other hand, in [18] the authors consider a space-time coding scheme for the point-to-point channel with imperfect channel knowledge. However, these existing works cannot be extended to robust transceiver design for the MIMO IC [21].

Researchers have utilized minimum mean square error criterion to design robust transceiver for the MIMO IC [22]. They improved robustness in presence of the channel uncertainty. Authors in [21] proposed a transceiver design that enforces robustness against imperfect CSI as well as providing a fair performance among the users in the interference channel. They adopted worst-case optimization approach to improve robustness and fairness.

*1.2. Contributions*

This paper focused on sum rate improvement of the Max-SINR algorithm under CSI error. In the first proposed scheme, each transceiver adjusts its filter by maximizing the expected value of the signal-to-interference-plus-noise ratio (SINR). Realized SINRs for different realizations of the CSI error matrices, are samples of the SINR random variable. In some cases, this random variable may have large variance. Therefore, the realized SINRs could be very far from the expected value. In the second proposed algorithm, each node adjusts its transmit/receive filter to minimize the variance. This design hedges against variability.

Two robust algorithms are designed based on the reciprocity of wireless propagation. There is a high correlation between the original and reciprocal channel's gains in communication systems working in a time division duplex (TDD) mode. Reciprocity has been exploited by researchers in [23-25]. Algorithms are implemented in a distributed manner with only local channel knowledge required. In other words, each user only needs to estimate the channel between its transmitter and receiver.

Monte Carlo simulations demonstrate that the first proposed algorithm, i.e. expectation maximization (EM), achieves higher sum rates compared to the existing algorithms in [21], and [22]. The second proposed algorithm, that is variation minimization (VM), provides a SINR with low variance. Moreover, VM helps to mitigate the effect of the CSI error, but not as satisfactory as EM scheme. Taylor series expansion is exploited to approximate the influence of CSI imperfection on the statistical properties, e. g. mean and variance. Numerical results show that more accurate approximation can be achieved with less error variance / signal power. For practical applications, when estimated mean / variance is used for maximization / minimization, the proposed transceivers will lead to sum rate improvement / SINR with low variance.

*1.3. Organization*

The rest of this paper is organized as follows. System model is studied in section 2. Sections 3 and 4 propose robust transceivers based on expectation maximization and variation minimization, respectively. Simulation results are presented in section 5 and concluding remarks are drawn in section 6.

## 2. System Model

In the system model under study, each node works in a TDD mode. At the first time slot, nodes on the left hand side send the data to the right hand side nodes, shown in Fig. 1. At the next time slot, nodes on the left hand side receive the data, indicating a change in the roles of the nodes. (This is clear in Fig. 2 following where the reciprocal network is described). The terms, original and reciprocal channels, are used to distinguish between two time slots. Hence, a transmitter in the reciprocal channel plays the role of an original network's receiver and vice versa.

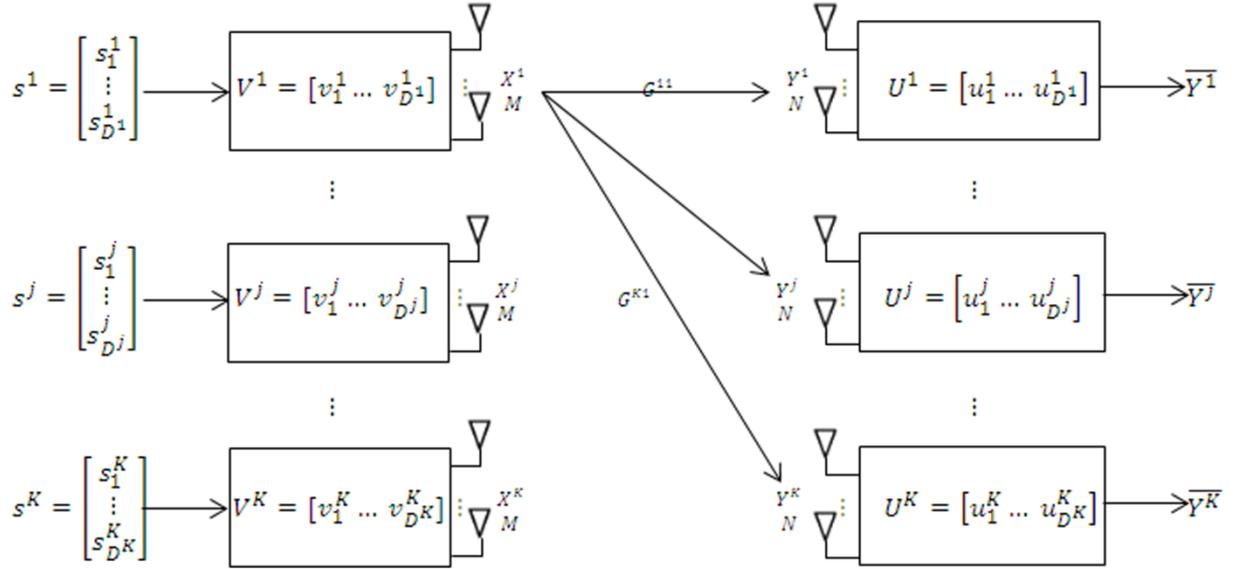

**Fig. 1.** *System model of the Original channel.*

MIMO IC with $K$ transmitter-receiver pairs is considered in this paper. The $j^{th} \in \mathcal{K}$ transmitter and the $k^{th} \in \mathcal{K} = \{1, \dots, K\}$ receiver are equipped with $M$ and $N$ antennas, respectively, as shown in Fig. 1. The $j^{th}$ transmitter sends $D^j$ independent data streams $s^j = [s_1^j \ ... \ s_{D^j}^j]^t$ to its intended receiver. Symbol vector has circularly symmetric complex Gaussian distribution with zero mean and covariance matrix $PI$, $s^j \sim CN(0, PI)$. True and estimated channel coefficients between transmitter $j$ and receiver $k$ are denoted by $G^{kj}$ and $H^{kj}$, respectively. In practical scenarios, there is a mismatch between true and estimated channels, as stated below:

$Error\ Model: G^{kj} = H^{kj} + E^{kj}$,

$Error\ Matrix: E^{kj}\ with\ i.i.d\ CN(0, \sigma^2)\ elements.$ (1)

The received signal at receiver $k$ is:

$Received\ Signal: Y^k = \sum_{j=1}^{K}(H^{kj} + E^{kj})X^j + Z^k$,

$Transmitted\ Signal\ Vector: X^j$, (2)

$AWGN\ Vector: Z^k \sim CN(0, N_0 I).$

In order to maximize system throughput, a beamforming strategy based on the interference alignment is used.

$Precoding\ Model: X^j = V^j s^j$,

$Precoder: M \times D^j\ matrix\ V^j = [v_1^j \ ... \ v_{D^j}^j]$, (3)

$Columns\ of\ Precoder: v_d^j\ with\ unit\ norms.$

Receiver $k$ decodes the transmitted symbol vector $s^k$ using the interference suppression matrix.

$Decoding\ Model: \overline{Y^k} = U^{k\dagger} Y^k$,

$Decoder: N \times D^k\ matrix\ U^k = [u_1^k \ ... \ u_{D^k}^k].$ (4)

For the *K*-user MIMO IC, reciprocal channel is obtained by switching the role of transmitters and receivers. True and estimated channels indicated by $\overleftarrow{G^{jk}}$ and $\overleftarrow{H^{jk}}$, are $M \times N$ matrices. Error matrix is denoted by $\overleftarrow{E^{jk}}$ with element distribution $CN(0, \sigma^2)$. The relation between the original and reciprocal channels' gains is $\overleftarrow{G^{jk}} = G^{kj\dagger}$. The $\overleftarrow{V^k}$ is precoder and $\overleftarrow{U^j}$ indicates receive interference suppression matrices on the reciprocal channel. Since the receivers of the reciprocal channel play the role of the original network's transmitters and vice versa, therefore $\overleftarrow{V^k} = U^k$ and $\overleftarrow{U^j} = V^j$. Fig. 2 shows the reciprocal network.

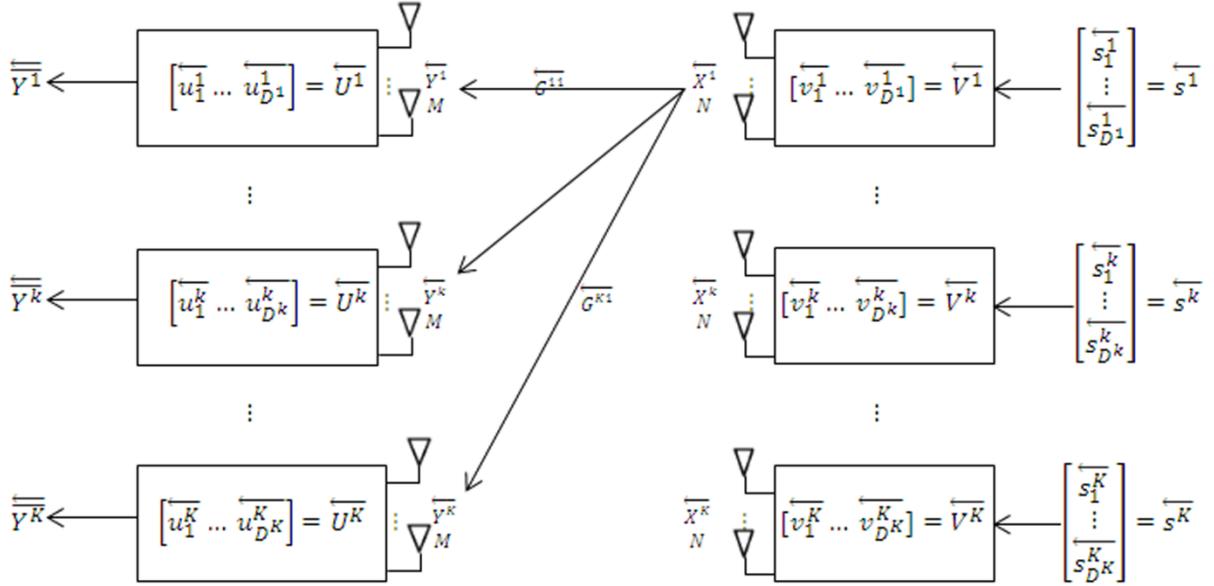

**Fig. 2.** *System model. Reciprocal network is obtained by switching the roles of transmitters and receivers in the original channel.*

According to the system model, the SINR value for the $d^{th}$ data stream at $k^{th}$ receiver is expressed by (where, $\|.\|$ denotes the Euclidian norm.)

$$SINR_d^k = \frac{P\left\|u_d^{k\dagger} G^{kk} v_d^k\right\|^2}{P\sum_{j=1}^{K}\sum_{m=1}^{D^j}\left\|u_d^{k\dagger} G^{kj} v_m^j\right\|^2 - P\left\|u_d^{k\dagger} G^{kk} v_d^k\right\|^2 + N_0\|u_d^k\|^2}. \qquad (5)$$

## 3. Robust Transceiver Design I

In this section, the first algorithm is formulated. The algorithm starts with arbitrary transmit and receive filters and then iteratively updates these filters to yield solution. The goal is to achieve robust transceiver by optimization problem $\max_{u_d^k} E[SINR]$. The iterative algorithm alternates between the original and reciprocal networks $\max_{\overleftarrow{u_d^j}} E[\overleftarrow{SINR}]$. Within each network, only the receivers update their filters.

In the following, first, approximate expression for the mean value is computed and then the optimization problem is solved.

### 3.1. Estimate the Mean of $SINR_d^k$

In (6) $E[SINR_d^k]$ is computed in terms of the function $SINR_d^k = \frac{num}{den}$ and the probability density function $f(num, den)$

$$E[SINR_d^k] = \int_{-\infty}^{\infty} \frac{num}{den} f(num, den) d.num \times d.den \,.^1 \tag{6}$$

Unfortunately, a closed form solution cannot be derived for the integration in (6). Hence, the approximate mean should be found. If $f(num, den)$ is concentrated near its mean, then estimation of the mean value can be expressed by

$$E[SINR_d^k|H^{kj}] \cong \frac{\mu_1}{\mu_2} = \frac{u_d^{k\dagger}[T_d^k + P\sigma^2 I]u_d^k}{u_d^{k\dagger}[S^k - T_d^k + (P\sigma^2 \sum_{j=1}^{K} D^j - P\sigma^2 + N_0)I]u_d^k}. \quad (7)$$

---

[1] The random variables, $num$ and $den$, are used to represent signal-to-interference-plus-noise ratio as a rational function $SINR_d^k = \frac{num}{den}$ and are defined by:
$num = P u_d^{k^\dagger} \left[ G^{kk} v_d^k v_d^{k^\dagger} G^{kk^\dagger} \right] u_d^k \,,$
$den = u_d^{k^\dagger} \left[ P \sum_{j=1}^{K} \sum_{m=1}^{D^j} \left( G^{kj} v_m^j v_m^{j^\dagger} G^{kj^\dagger} \right) - P \left( G^{kk} v_d^k v_d^{k^\dagger} G^{kk^\dagger} \right) + N_0 I \right] u_d^k \,.$

Where $S^k = P \sum_{j=1}^{K} \sum_{m=1}^{D^j} H^{kj} v_m^j v_m^{j\dagger} H^{kj\dagger}$ and $T_d^k = P H^{kk} v_d^k v_d^{k\dagger} H^{kk\dagger}$ denote, respectively, the estimated covariance matrix of all data streams observed by the receiver $k$ and estimated covariance matrix of the $d^{th}$ desirable data stream. Since the estimation of mean is common in two algorithms, it is provided in appendix A.

*3.2. Iterative Solution*

To obtain columns of $U^k$, the derivative of (7) with respect to $u_d^k$ should be obtained and then set equal to zero. Thus, $u_d^k$ should satisfy the following vector equation (i.e. the derivative of the numerator multiplied by $\mu_2$ should be equal to the derivative of denominator multiplied by $\mu_1$).

$$\mu_2 [T_d^k + P\sigma^2 I] u_d^k = \mu_1 [S^k - T_d^k + (P\sigma^2 \sum_{j=1}^{K} D^j - P\sigma^2 + N_0)I] u_d^k. \quad (8)$$

$\mu_1 = E[num|H^{kj}] = u_d^{k\dagger}[T_d^k + P\sigma^2 I]u_d^k$. Eq. 31 in appendix A

$\mu_2 = E[den|H^{kj}] = u_d^{k\dagger}[S^k - T_d^k + (P\sigma^2 \sum_{j=1}^{K} D^j - P\sigma^2 + N_0)I]u_d^k$. Eq. 32 in appendix A

The above vector equation is rearranged as follow by moving the terms involving $S^k u_d^k$ and $u_d^k$ to left and $T_d^k u_d^k$ to the right hand side: (where, the scalar coefficient is $\Omega_d^k$.)

$$\mu_1(S^k u_d^k + \Omega_d^k u_d^k) = (\mu_1 + \mu_2) T_d^k u_d^k, \quad (9)$$

$$\Omega_d^k = P\sigma^2 \sum_{j=1}^{K} D^j - P\sigma^2 \frac{\mu_1+\mu_2}{\mu_1} + N_0 \ . \tag{10}$$

According to the definition, $T_d^k u_d^k$ is the product of scalar value $v_d^{k\dagger} H^{kk\dagger} u_d^k$ and vector $H^{kk} v_d^k$. It is concluded that $T_d^k u_d^k$ is in the direction of $H^{kk} v_d^k$. Furthermore, only the directions are important. Hence, the scalar factors $\mu_1$, $(\mu_1 + \mu_2) v_d^{k\dagger} H^{kk\dagger} u_d^k$ can be removed from (9). Then, the unit vector maximizes (7) is given by

$$u_d^k = \frac{(S^k + \Omega_d^k I)^{-1} H^{kk} v_d^k}{\|(S^k + \Omega_d^k I)^{-1} H^{kk} v_d^k\|} \ . \tag{11}$$

Now, consider the reciprocal network. The transmit precoding matrices, $\overleftarrow{V^k}$, are the receive interference suppression matrices $U^k$ from the original network, whose columns are given by (11). The optimal $d^{th}$ unit column of $\overleftarrow{U^j}$ is given by

$$\overleftarrow{u_d^j} = \frac{(\overleftarrow{S^j} + \overleftarrow{\Omega_d^j} I)^{-1} \overleftarrow{H^{jj}} \overleftarrow{v_d^j}}{\|(\overleftarrow{S^j} + \overleftarrow{\Omega_d^j} I)^{-1} \overleftarrow{H^{jj}} \overleftarrow{v_d^j}\|} \ . \tag{12}$$

Now, the receive interference suppression matrices in the reciprocal network replace $V^j \ \forall j \in \mathcal{K}$, and then $U^k \ \forall k \in \mathcal{K}$ are updated based on them. It is seen from (10) and (11), that $\Omega_d^k$ and $u_d^k$ are interdependent. Therefore, prior to repeat steps, $\Omega_d^k$ should be computed at the end of each iteration. To summarize the iterative procedure the steps are given in Fig. 3.

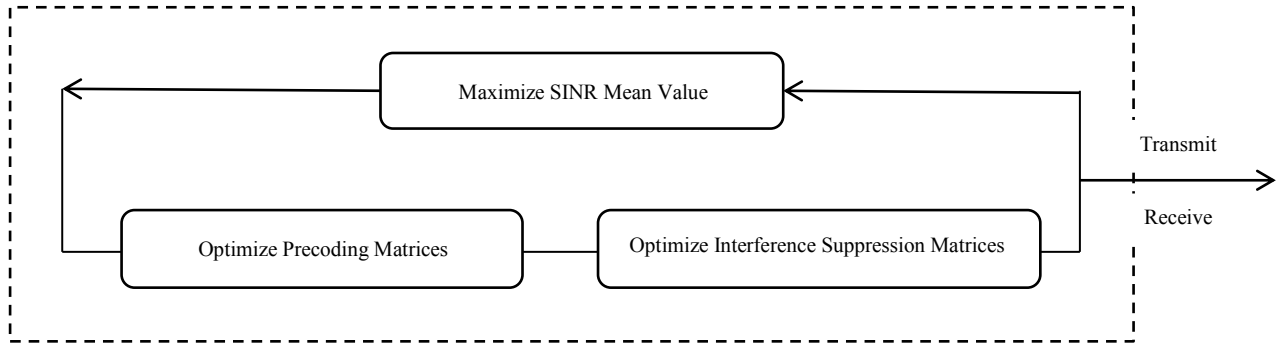

**(a)**

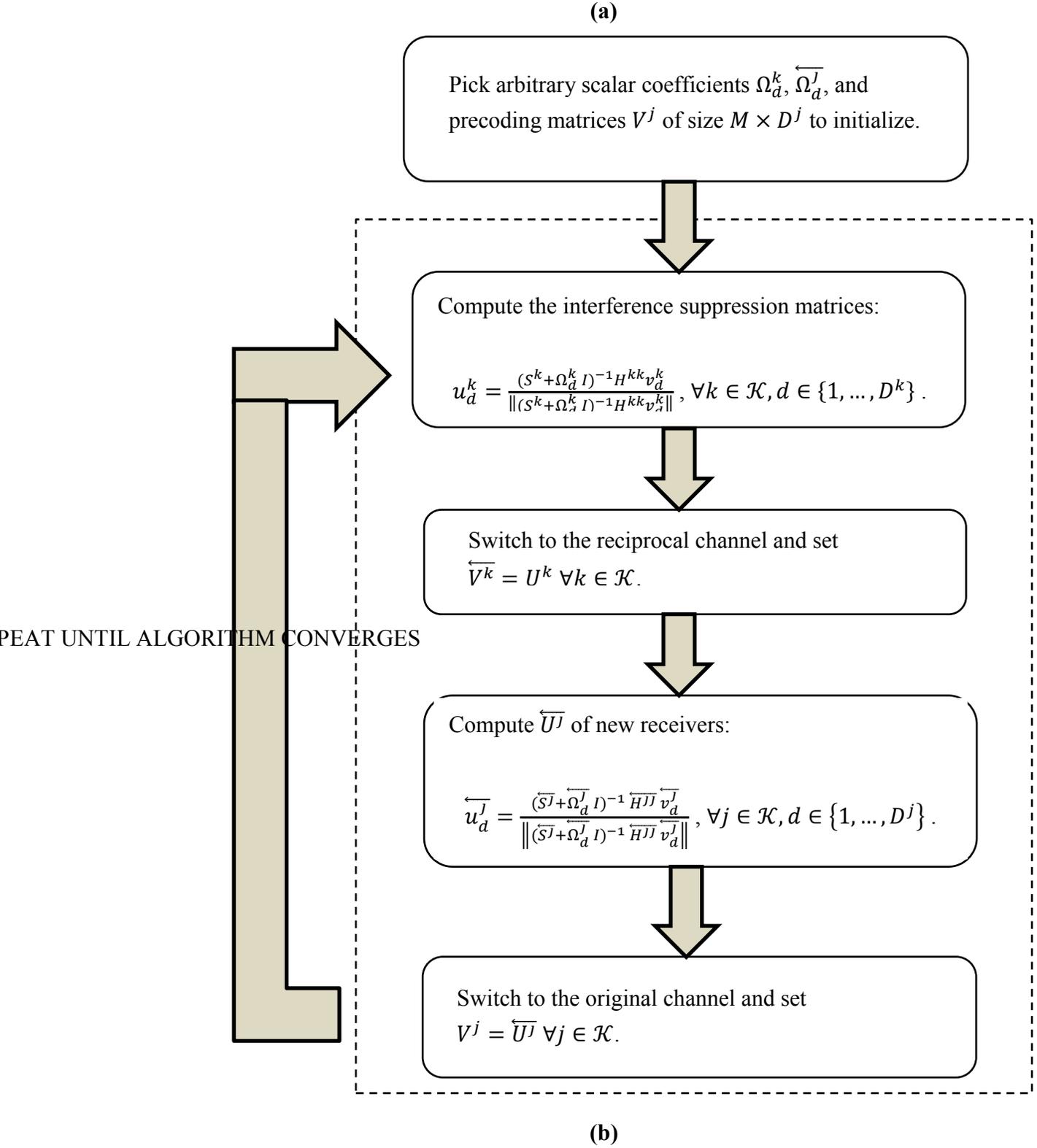

**(b)**

*Fig. 3*. **(a)** *Schematic view of robust transceiver design I.* **(b)** *Algorithm: Robust Distributed Transceiver Design.*

It can be proved that EM filters for the special case $\sigma^2 = 0$, are transmit / receive matrices of the Max-SINR algorithm (Proof is provided in appendix B.).

In order to implement the algorithm in a distributed manner, receiver $k$ needs to know about $H^{kk}$ and $S^k$ which are locally available. The covariance matrix $S^k$ can be estimated from the autocorrelation of the received signal $Y^k$. Substituting $X^j = \sum_{d=1}^{D^j} v_d^j s_d^j$ into (2) yields

$$E\left[Y^k Y^{k\dagger}\right] = S^k + \left(P\sigma^2 \sum_{j=1}^{K} D^j + N_0\right)I, \tag{13}$$

where, the expectation is computed over error and noise. The receiver $j$ in the reciprocal channel can learn $\overleftarrow{U^j}$ in a similar manner. For TDD systems, the transmitters can estimate the channels from the sounding signals received in the reverse link [21, section II-A]. Using MMSE channel prediction, the CSI estimate $H^{kk}$ is obtained, whereas the CSI error $E^{kj}$ is Gaussian distributed and independent from the CSI estimate $H^{kj}$ [21, section II-A].

### 3.3. Proof of Convergence

Now, the convergence of the proposed EM algorithm is demonstrated. Equivalent problem is considered. EM can be written as follow

$$\max \frac{u_d^{k\dagger} Q u_d^k}{u_d^{k\dagger} F u_d^k}, \tag{14}$$

where, $Q = Q^\dagger = T_d^k + P\sigma^2 I \geq 0$, and $F = F^\dagger = S^k - T_d^k + \left(P\sigma^2 \sum_{j=1}^{K} D^j - P\sigma^2 + N_0\right)I > 0$ are matrices and $u_d^k$ indicates optimization variable. It is shown in [26] that (14) is equivalent to

$$\max u_d^{k\dagger} Q u_d^k, \tag{15}$$
$$\text{s. t. } u_d^{k\dagger} F u_d^k = 1.$$

For the equivalent problem, the Lagrangian function is given by $l(u_d^k, \lambda) = u_d^{k\dagger} Q u_d^k + \lambda\left(1 - u_d^{k\dagger} F u_d^k\right)$. The solution $u_d^{k*}$ is the eigenvector corresponding to the maximal eigenvalue of $F^{-1}Q$, and the Lagrange multiplier is $\lambda^* = u_d^{k*\dagger} Q u_d^{k*}$. ∎

The metric is defined in (16). It is proved here that each step in the algorithm increases the metric. Since it cannot increase unboundedly, this implies that equivalent problem converges and consequently algorithm Fig 3 also converges. It is important to note that the metric is the same for both original and reciprocal networks.

$$\max_{\substack{V^j \text{ and } U^K \\ \forall\, j \text{ and } k \in \mathcal{K}}} metric = \sum_{k=1}^{K} \sum_{d=1}^{D^k} l(u_d^k, \lambda). \tag{16}$$

Accordingly:

$$\max_{\substack{U^K \\ \forall\, k \in \mathcal{K}}} metric = \sum_{k=1}^{K} \sum_{d=1}^{D^k} \max_{u_d^k} l(u_d^k, \lambda). \tag{17}$$

In other words, given $V^j\ \forall\, j \in \mathcal{K}$, Step 1 increases the value of (16) over all possible choices of $U^k\ \forall\, k \in \mathcal{K}$. The filter $\overleftarrow{U^J}$ computed in Step 3, based on $\overleftarrow{V^k} = U^k$, also maximizes the metric in the reciprocal channel (18).

$$\max_{\substack{\overleftarrow{U^J} \\ \forall\, j \in \mathcal{K}}} \overleftarrow{metric},$$

$$\overleftarrow{metric} = \sum_{j=1}^{K} \sum_{d=1}^{D^j} \overleftarrow{l}\left(\overleftarrow{u_d^J}, \lambda\right) = \sum_{j=1}^{K} \sum_{d=1}^{D^j} \overleftarrow{u_d^J}^{\dagger} \overleftarrow{Q} \overleftarrow{u_d^J} + \lambda\left(1 - \overleftarrow{u_d^J}^{\dagger} \overleftarrow{F} \overleftarrow{u_d^J}\right). \tag{18}$$

Since $\overleftarrow{V^k} = U^k$ and $\overleftarrow{U^J} = V^j$, the metric remains unchanged in the original and reciprocal networks, according to following equation:

$$\overline{metric} = \sum_{j=1}^{K}\sum_{d=1}^{D^j} u_d^{j\dagger}[T_d^j + P\sigma^2 I]u_d^j +$$

$$\sum_{j=1}^{K}\sum_{d=1}^{D^j} \lambda_d^j \left(1 + u_d^{j\dagger}[T_d^j - (P\sigma^2 \sum_{k=1}^{K} D^k - P\sigma^2 + N_0)I]u_d^j\right) - \quad (19)$$

$$P\sum_{j=1}^{K}\sum_{d=1}^{D^j}\sum_{k=1}^{K}\sum_{m=1}^{D^k} \lambda_d^j u_m^{k\dagger} H^{kj} v_d^j v_d^{j\dagger} H^{kj\dagger} u_m^k = metric.$$

Therefore, Step 3 also can increase the value of (16). Since the value of (16) is monotonically increased after every iteration, convergence of the algorithm is guaranteed.

## 4. Robust Transceiver Design II

Realized SINRs for different realizations of the CSI error matrices, are samples of the SINR random variable. This random variable can has large variance. Hence, the realized SINR, depending on the particular realization of the CSI error matrix, could be very far from the expected value. Therefore, to hedge against such variability, each receiver adjusts its receive interference suppression matrix based on minimizing SINR variance [27]:

$$\min_{u_d^k} VAR[SINR_{lb\,d}^{\,k}], \forall d \in \{1,\dots,D^k\}. \quad (20)$$

According to $VAR(x) = E(x^2) - E(x)^2$, the $VAR(x)$ is minimized by minimizing $E(x^2)$ and maximizing $E(x)^2$. These two terms may not attain their best values for the same transceivers in some MIMO IC system model, due to the contradiction between them. It will be shown that the VM scheme presents sum data rate (maximize $E(x)$) lower than the Max-SINR algorithm but it enables transceivers to hedge against variability for the primary scenario in simulation results. For the second scenario, the VM improves sum data rate superior to the Max-SINR and provides SINR with low variance, simultaneously.

In section 4.1, the value of $VAR\left[SINR_{lb_d}^k\right]$ is approximated by using statistical linearization argument. The iterative solution is similar to Algorithm I.

## 4.1. Estimating the Variance of $SINR_{lb_d}^k$

Lower bound on the SINR is derived in terms of norms of error matrices in [21]. Lower bound is

$$SINR_{lb_d}^k = \frac{P\left\|u_d^{k\dagger}H^{kk}v_d^k\right\|^2 - Pe^{kk}\left\|u_d^k\right\|^2}{P\sum_{j=1}^{K}\sum_{m=1}^{D^j}\left\|u_d^{k\dagger}H^{kj}v_m^j\right\|^2 + P\left\|u_d^k\right\|^2\sum_{j=1}^{K}e^{kj}D^j - P\left\|u_d^{k\dagger}H^{kk}v_d^k\right\|^2 - Pe^{kk}\left\|u_d^k\right\|^2 + N_0\left\|u_d^k\right\|^2}. \quad (21)$$

$\left\|E^{kj}\right\|^2$ is the Euclidian norm of error matrix between transmitter $j$ and receiver $k$, denotes by $e^{kj}$ in equation (21). $\frac{\left\|E^{kj}\right\|^2}{\sigma^2/2}$, is the sum of the second power of $2NM$ real independent Gaussians, each having a unit variance. Therefore, it has a Chi-square distribution with $2NM$ degrees of freedom, $\chi^2_{2NM}$. Hence, one can conclude $E[e^{kj}] = NM\sigma^2$ and $VAR[e^{kj}] = NM\sigma^4$ [28].

Equation (21), $SINR_{lb_d}^k$, is a function of random vector $e^k = [e^{k1} \quad ... \quad e^{kK}]^t$. It is clear from covariance matrix, $Cov(e^k) = \begin{bmatrix} MN\sigma^4 & ... & 0 \\ \vdots & \ddots & \vdots \\ 0 & ... & MN\sigma^4 \end{bmatrix}$, that the variance of each element is sufficiently small for practical applications. Besides, the covariance between each two components is zero. It is concluded that the PDF of $e^k$ is concentrated near its mean and it is negligible outside a neighborhood around the mean value. Again by using the statistical linearization argument, first order Taylor series expansion of $SINR_{lb_d}^k$ around the mean value

$\theta = E[e^k] = [MN\sigma^2 \quad \ldots \quad MN\sigma^2]^t$ is employed to yield:

$$SINR_{lb_d}^{\ k} \cong SINR_{lb_d}^{\ k}(\theta) + \frac{\partial SINR_{lb_d}^{\ k}(\theta)}{\partial e^{k1}}(e^{k1} - NM\sigma^2) + \cdots + \frac{\partial SINR_{lb_d}^{\ k}(\theta)}{\partial e^{kK}}(e^{kK} - NM\sigma^2)$$

$$= SINR_{lb_d}^{\ k}(\theta) + \frac{\partial SINR_{lb_d}^{\ k}(\theta)^t}{\partial e^k}(e^k - \theta) ,^2 \qquad (22)$$

Using (22) for approximating the variance, following equation is obtained.

$$VAR[SINR_{lb_d}^{\ k}] \cong E\left[\left(SINR_{lb_d}^{\ k}(\theta) + \frac{\partial SINR_{lb_d}^{\ k}(\theta)^t}{\partial e^k}(e^k - \theta) - E[SINR_{lb_d}^{\ k}]\right)^2\right]. \qquad (23)$$

Since exact computation of $E[SINR_{lb_d}^{\ k}]$ is not feasible, approximate mean $SINR_{lb_d}^{\ k}(\theta)$ is used (Estimation of mean is provided in appendix A.). Therefore, estimation of variance leads to

$$VAR[SINR_{lb_d}^{\ k}] \cong \frac{\partial SINR_{lb_d}^{\ k}(\theta)^t}{\partial e^k} Cov(e^k) \frac{\partial SINR_{lb_d}^{\ k}(\theta)}{\partial e^k} .^3 \qquad (24)$$

The elements of $\frac{\partial SINR_{lb_d}^{\ k}}{\partial e^k}$ are given by

$$\frac{\partial SINR_{lb_d}^{\ k}(\theta)}{\partial e^{kj}} = -\frac{PD^j\left(u_d^{k\dagger}u_d^k\right)\left(u_d^{k\dagger}[T_d^k - PMN\sigma^2 I]u_d^k\right)}{\left(u_d^{k\dagger}\left[S^k - T_d^k + \left(PMN\sigma^2 \sum_{j=1}^K D^j - PMN\sigma^2 + N_0\right)I\right]u_d^k\right)^2}, \forall j \in \mathcal{K}, j \neq k ,$$

$$\frac{\partial SINR_{lb_d}^{\ k}(\theta)}{\partial e^{kk}} = -\frac{P\left(u_d^{k\dagger}u_d^k\right)\left(u_d^{k\dagger}\left[S^k + (D^k - 2)T_d^k + \left(PMN\sigma^2 \sum_{\substack{j=1 \\ j \neq k}}^K D^j + N_0\right)I\right]u_d^k\right)}{\left(u_d^{k\dagger}\left[S^k - T_d^k + (PMN\sigma^2 \sum_{j=1}^K D^j - PMN\sigma^2 + N_0)I\right]u_d^k\right)^2}, j = k . \qquad (25)$$

According to (24) and (25), the estimated variance is

$$VAR[SINR_{lb_d}^{\ k}|H^{kj}] \cong$$

---

[2] The $\frac{\partial SINR_{lb_d}^{\ k}}{\partial e^k}$ is a $K \times 1$ random vector whose $j^{th}$ component is the derivative of $SINR_{lb_d}^{\ k}$ with respect to $e^{kj}$.

[3] In estimating variance both mean vector and covariance matrix are required [29]; hence, lower bound on SINR is chosen because covariance matrix of random vector is known.

$$MNP^2\sigma^4(u_d^{k\dagger}u_d^k)^2 \frac{(u_d^{k\dagger}\left[S^k+(D^k-2)T_d^k+(PMN\sigma^2\sum_{\substack{j=1\\j\neq k}}^{K} D^j+N_0)I\right]u_d^k)^2 + \left(\sum_{\substack{j=1\\j\neq k}}^{K}(D^j)^2\right)\left(u_d^{k\dagger}[T_d^k-PMN\sigma^2 I]u_d^k\right)^2}{\left(u_d^{k\dagger}\left[S^k-T_d^k+(PMN\sigma^2\sum_{j=1}^{K}D^j-PMN\sigma^2+N_0)I\right]u_d^k\right)^4} \quad (26)$$

## 4.2. Iterative Solution

Briefly, $u_d^k$ should satisfy the following vector equation: To obtain columns of $U^k$, the equation $\frac{\partial VAR[SINR_{lbd}^k]}{\partial u_d^k}=0$ should be solved. Thus, $u_d^k$ should satisfy the following vector equation (The terms involving $S^k u_d^k$ and $u_d^k$ are moved to left and the term $T_d^k u_d^k$ is moved to the right hand side).

$$\alpha_d^k S^k u_d^k + \beta_d^k u_d^k = \zeta_d^k T_d^k u_d^k. \quad (27)$$

In the above vector equation, $\alpha_d^k$, $\beta_d^k$, and $\zeta_d^k$ denote scalar coefficients and are expanded as follow

$$\alpha_d^k = 2ac - 4a^2 - 4\left(\sum_{\substack{j=1\\j\neq k}}^{K}(D^j)^2\right)b^2;$$

$$\beta_d^k = 2a^2c + 2\left(\sum_{\substack{j=1\\j\neq k}}^{K}(D^j)^2\right)b^2c + 2\left(PMN\sigma^2\sum_{\substack{j=1\\j\neq k}}^{K} D^j + N_0\right)ac - 2\left(\sum_{\substack{j=1\\j\neq k}}^{K}(D^j)^2\right)\times$$
$$PMN\sigma^2 bc - 4\left(a^2 + \left(\sum_{\substack{j=1\\j\neq k}}^{K}(D^j)^2\right)b^2\right)\left(PMN\sigma^2\sum_{j=1}^{K} D^j - PMN\sigma^2 + N_0\right); \quad (28)$$

$$\zeta_d^k = -\left(2(D^k-2)ac + 2\left(\sum_{\substack{j=1\\j\neq k}}^{K}(D^j)^2\right)bc + 4a^2 + 4\left(\sum_{\substack{j=1\\j\neq k}}^{K}(D^j)^2\right)b^2\right).$$

In (28), the parameters of the scalar coefficients ($a$, $b$, and $c$) should be substituted by:

$$a = u_d^{k\dagger}\left[S^k + (D^k - 2)T_d^k + (PMN\sigma^2 \sum_{\substack{j=1 \\ j \neq k}}^{K} D^j + N_0)I\right]u_d^k \ ;$$

$$b = u_d^{k\dagger}[T_d^k - PMN\sigma^2 I]u_d^k \ ; \qquad (29)$$

$$c = u_d^{k\dagger}[S^k - T_d^k + (PMN\sigma^2 \sum_{j=1}^{K} D^j - PMN\sigma^2 + N_0)I]u_d^k \ .$$

The unit vector that minimizes (26) is given by

$$u_d^k = \frac{(S^k + \Psi_d^k I)^{-1} H^{kk} v_d^k}{\|(S^k + \Psi_d^k I)^{-1} H^{kk} v_d^k\|}, \quad \Psi_d^k = \frac{\beta_d^k}{\alpha_d^k} \ . \qquad (30)$$

The iterative procedure, Fig. 4, is algorithmically identical to the Algorithm I. Moreover, the distributed implementation explained previously can be applied again.

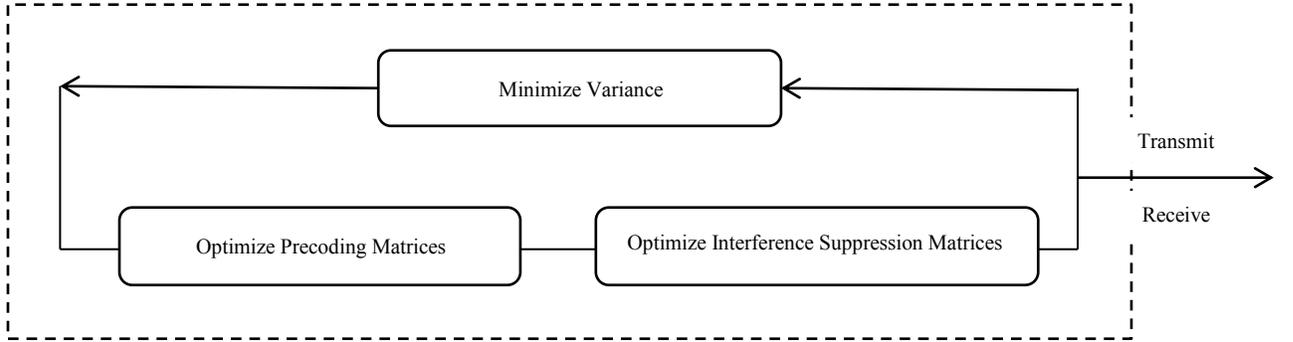

*Fig. 4. Schematic view of Algorithm II.*

## 5. Simulation Results

In this section, simulation results for the sum data rate and variance of SINR are presented. Also, the influence of the mean value approximation, and other influential parameters on the accuracy of the approximation are determined via Monte Carlo simulations. The proposed robust transceiver design algorithms are evaluated through comparison to the following algorithms:

1. Leakage interference minimization [13]

2. Max-SINR algorithm [13]

3. Minimum Mean Square Error (MMSE) [15]

4. Robust MMSE [22]

5. Worst-case optimization approach [21].

The use of [13], [15] for comparison is not actually fair because these papers design filter with perfect CSI. The reason is that, their sum-rates do not increase linearly with SNR, and they achieve lower sum-rates compared to our proposed algorithm. (Mean loss in sum rate compared to perfect CSI is shown in Fig. 7 e.g. for Max-SINR algorithm.) [21], [22] design robust transceiver for MIMO interference network, under imperfect CSI.

The channel is modeled as Rayleigh fading. Channel coefficients $\left[G^{kj}\right]_{nm}$ are i.i.d. zero mean unit variance circularly symmetric complex Gaussian. MIMO IC has $K=4$ users with $N=M=3$ antennas at transmitters and receivers and each user transmits $d=1$ data stream denoted by $(3\times 3,1)^4$. Simulation results are also presented for $(4\times 4,2)^3$ MIMO IC.

20 error matrices are generated for a true channel and numerical results are averaged over them. Averaging process over the error is repeated for 20 true channels to eliminate the dependence of the numerical results on the true channel, randomly created; In other words the results are obtained after 400 Monte Carlo simulations. The stopping criterion for the convergence of the proposed iterative algorithms is 100 iterations. All numerical results are based on the SINR associated with the imperfect CSI[4].

## 5.1. Throughput Enhancement

Fig. 5 represents the sum rate comparison between the proposed and basic algorithms for $(3 \times 3, 1)^4$ MIMO IC. The filters are designed with the error variance of $\sigma^2 = 0.1$. It can be observed that the EM scheme achieves higher sum rates compared to all the other schemes over the entire considered SNR[5] range. Proposed EM scheme achieves 7dB SNR gain over the Max-SINR algorithm at providing 14 b/s/Hz sum data rate and etc. Though it does not seem the VM algorithm improves the overall sum data rate as satisfactory as the EM scheme, but it can be roughly concluded that the VM algorithm achieves data rate as much as (slightly lower than) Max-SINR. Sum rate of algorithm 5 is shown in Fig. 5. It presents sum data rate performance as robust MMSE.

---

[4] Average data rate is defined as the average throughput (i.e. the bits/s/Hz successfully delivered to the receiver). Specifically, the throughput of $d^{th}$ data stream at $k^{th}$ receiver is given by $R_d^k$ ($R_d^k \leq C_d^k$), where $R_d^k = \log(1 + sinr_d^k)$ and

$$sinr_d^k = \frac{P\|u_d^{k\dagger}H^{kk}v_d^k\|^2}{P\sum_{j=1}^{K}\sum_{m=1}^{D^j}\|u_d^{k\dagger}H^{kj}v_m^j\|^2 - P\|u_d^{k\dagger}H^{kk}v_d^k\|^2 + N_0\|u_d^k\|^2},$$

and $C_d^k = \log(1 + SINR_d^k)$ is the actual instantaneous mutual information. The overall sum rate of the system is given by $R = \sum_{k=1}^{K}\sum_{d=1}^{D^k} R_d^k$.

[5] $\frac{P}{N_0}$ is SNR in the network, since all data streams are of power $P$ and $N_0$ is noise power at all receivers.

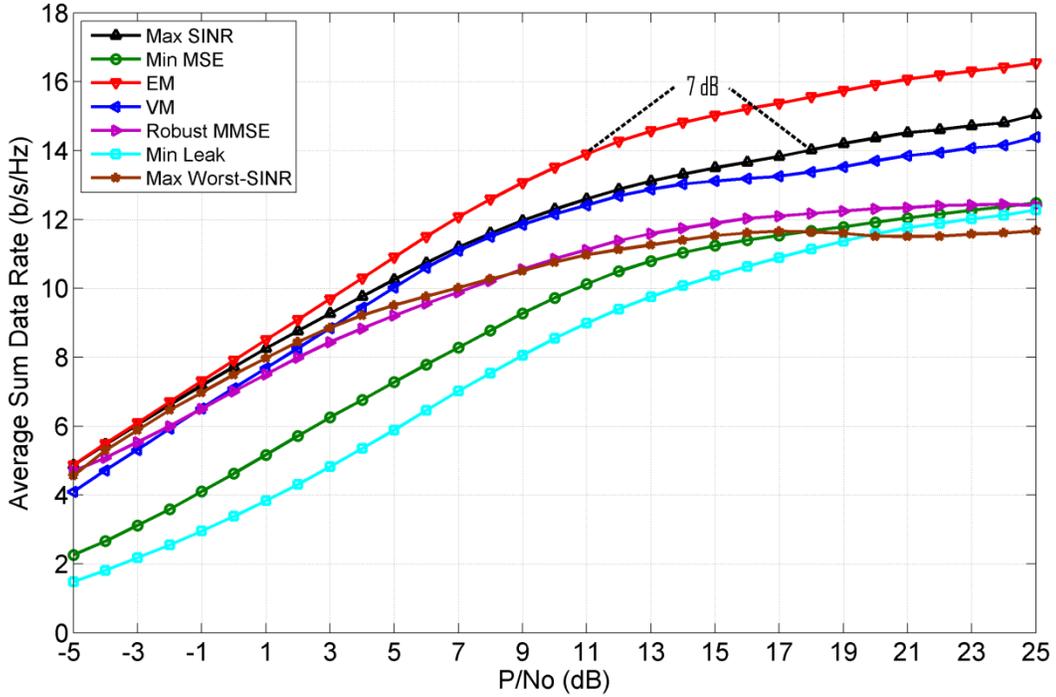

***Fig. 5.*** *Average sum data rate versus SNR. $K = 4$, $N = M = 3$, $d = 1$, $\sigma^2 = 0.1$.*

Fig. 6 shows sum rate for $(4 \times 4, 2)^3$ MIMO IC. Again, the EM scheme achieves higher sum rates compared to all the other schemes. In comparison with Max-SINR, it cannot achieve data rate higher than 12 b/s/Hz, while the EM scheme improves data rate up to 16 b/s/Hz. Compared to the Max-SINR, the VM helps to mitigate the effect of the CSI error more effectively, as shown in Fig. 6. The algorithm 5 achieves a data rate only superior to the Leakage interference minimization. It should be noted that, this scheme guaranties to provide better worst-case data rate.

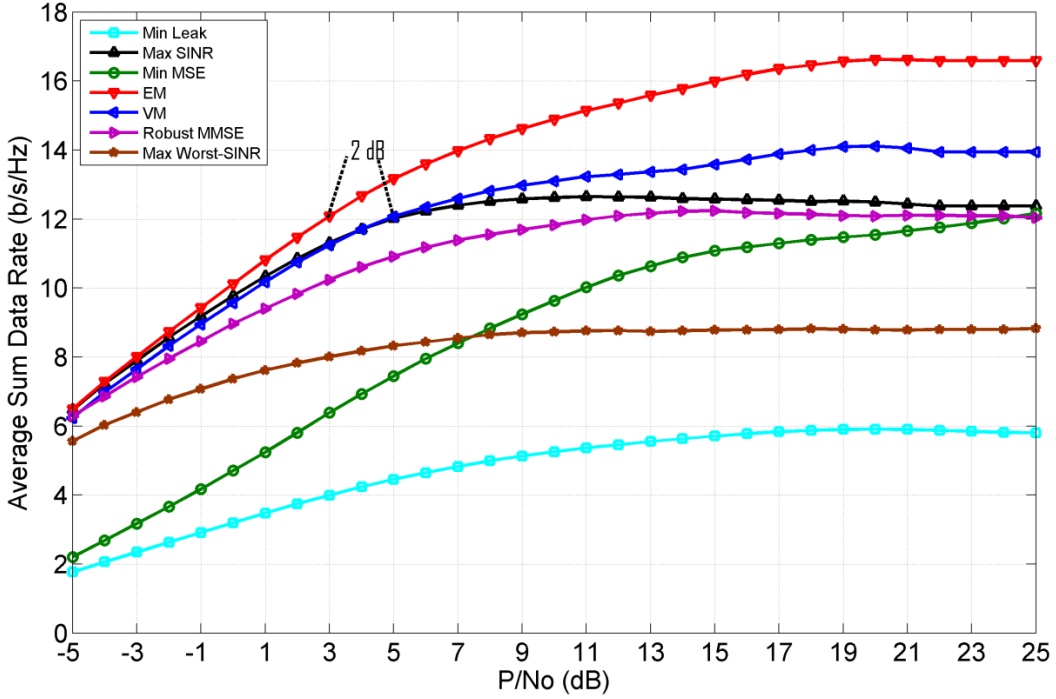

*Fig. 6.* *Average sum data rate versus SNR.* $K = 3$, $N = M = 4$, $d = 2$, $\sigma^2 = 0.1$.

The better performances of the proposed schemes compared to other algorithms are accountable to suitably: 1- Both proposed schemes improve the resilience against SINR degradation due to the CSI error. 2- Approximation (When approximate mean or variance is used for maximization or minimization, the proposed transceivers will lead to sum rate improvement).

## 5.2. Hedge against Variability

Average SINR variance in $(3 \times 3,1)^4$ and $(4 \times 4,2)^3$ MIMO ICs are reported in Table I and Table II, respectively. It can be observed that proposed VM scheme achieves lower variance compared to other schemes, when the amount of SNR (>10dB) is significant. In case of small SNR (<10dB), the algorithms have variance close to each other.

By Considering Fig. 5 and Table I jointly, it is concluded VM cannot achieve data rate higher than and SINR variance lower than Max-SINR simultaneously. This is due to the contradiction between minimizing $E(x^2)$ and maximizing $E(x)^2$ in VM algorithm.

Table I

Average SINR variance in $(3 \times 3, 1)^4$ interference network with error variance $\sigma^2 = 0.1$

| | | SNR (dB) | | | | | | | | |
|---|---|---|---|---|---|---|---|---|---|---|
| | Algorithm | 2 | 5 | 8 | 12 | 14 | 16 | 18 | 20 | 22 | 24 |
| $VAR_{error}[sinr_d^k|H^{kj}]$ | Max-SINR | 3.8 | 11.6 | 32.3 | 120.2 | 216.9 | 375.5 | 599.1 | 998.6 | 1593.3 | 2285.3 |
| | EM | 4.0 | 13.6 | 43.0 | 153.3 | 273.8 | 438.7 | 642.8 | 963.6 | 1429.9 | 2112.4 |
| | VM | 5.0 | 14.6 | 37.4 | 102.1 | 151.6 | 203.6 | 261.3 | 325.9 | 403.6 | 428.7 |

The comparative improvement in SINR variance level becomes negligible in $(4 \times 4, 2)^3$ MIMO IC, since variance of the VM approaches other algorithms, as presented in Table II. On the other hand, VM achieves data rate better than Max-SINR as seen in Fig. 6. Therefore, VM presents a balance between minimizing $E(x^2)$ and maximizing $E(x)^2$ in this scenario.

**Table II**

Average SINR variance in $(4 \times 4, 2)^3$ interference network with $\sigma^2 = 0.1$

| | | SNR (dB) | | | | | | | | |
|---|---|---|---|---|---|---|---|---|---|---|
| | Algorithm | 2 | 6 | 10 | 12 | 14 | 16 | 18 | 20 | 22 | 24 |
| $VAR_{error}[sinr_d^k|H^{kj}]$ | Max-SINR | 2.4 | 5.6 | 10.8 | 13.7 | 16.7 | 19.0 | 22.4 | 27.3 | 32.3 | 33.7 |
| | EM | 2.9 | 8.6 | 19.4 | 25.5 | 32.0 | 39.6 | 49.7 | 58.9 | 65.9 | 72.1 |
| | VM | 3.4 | 6.6 | 10.6 | 12.9 | 15.6 | 17.2 | 17.3 | 17.1 | 17.0 | 16.0 |

The cost of the proposed robust design methods compared with the contrast schemes is the complexity since the MMSE [15, equation 11], and Robust MMSE [22, equation 4] need the inverse operation of an N-by-N matrix only once to update $V^k$ (or $U^k$) at each user per iteration, whereas the proposed algorithms (equations 11 and 30) require d times. In the SINR maximizing algorithm [13, equation 30], contrast scheme 5 [21, equation 13], the transmit and receive filters are column-wise updated, as complex as the proposed algorithms.

*5.3. Conformity with Performance Evaluation in [4]*

Based on equation 21 in [4], it is expected that mean loss in sum rate compared to perfect CSI case increases unboundedly as SNR increases. In addition, based on equation 22 in [4], the achievable multiplexing gain should be equal to zero. These are confirmed in Fig. 7, where for the larger SNR, the wider gap between curves representing perfect CSI and imperfect CSI can be noticed. Slope of the curves or multiplexing gain when $\sigma^2 = 0.1$ is zero, too.

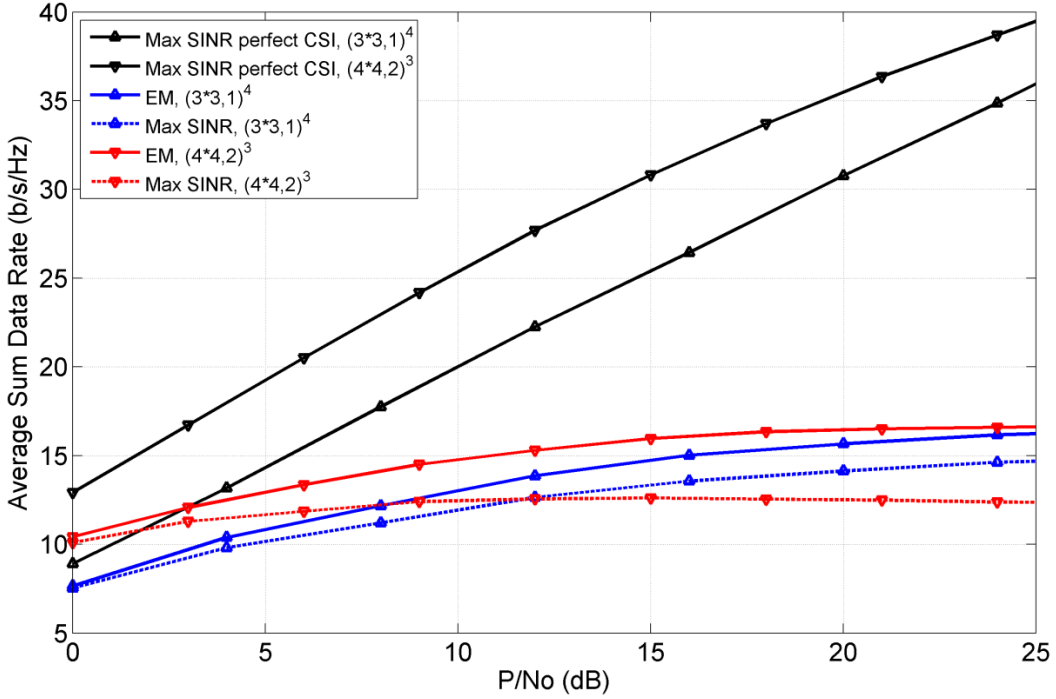

**Fig. 7.** *Average sum rates achieved by the EM and Max-SINR algorithms for $(3 \times 3, 1)^4$ and $(4 \times 4, 2)^3$ interference networks with $\sigma^2 = 0.1$.*

## 5.4. Accuracy of Approximation

It is straightforward to say as $\sigma^2$ decreases, the impact of any error in approximating average mutual information diminishes. On the other hand, signal power *P* scales with $\sigma^2$ as it is obvious from (7). Therefore, any approximation error will be attenuated as *P* decreases. But, here the impact of loss due to the mean approximation and other influential parameters on the accuracy of approximation are studied via Monte Carlo simulations. The EM algorithm is used by $(3 \times 3,1)^4$ MIMO IC to compute precoding and interference suppression matrices. The filters are designed with two CSI error variances $\sigma_1^2 = 0.05$ and $\sigma_2^2 = 0.1$. The numerical and theoretical values of average mutual information[6] of the MIMO IC versus SNR are depicted in Fig. 8.

For $\sigma_1^2 = 0.05$, the proposed approximation is within 15% and 18% of the true value when $SNR \leq 10dB$ and $SNR \leq 14dB$. For $\sigma_2^2 = 0.1$ similar statement stands when $SNR \leq 7dB$ and $SNR \leq 9dB$. The proposed approximation is within 22.16% and 24.81% of the true value, respectively for $\sigma_1^2$ and $\sigma_2^2$, over the entire considered SNR range. Therefore, by decreasing SNR and $\sigma^2$ the theoretical approximation will approach the true value of average mutual information.

---

[6] Theoretical or approximate capacity curve are found by substituting (7) into $C = \sum_{k=1}^{K} \sum_{d=1}^{D^k} \log(1 + SINR_d^k)$.

$$E[C|H^{kj}] \cong \sum_{k=1}^{K} \sum_{d=1}^{D^k} \log\left(1 + \frac{u_d^{k\dagger}\left[T_d^k + P\sigma^2 I\right]u_d^k}{u_d^{k\dagger}\left[S^k - T_d^k + \left(P\sigma^2 \sum_{j=1}^{K} D^j - P\sigma^2 + N_0\right)I\right]u_d^k}\right).$$

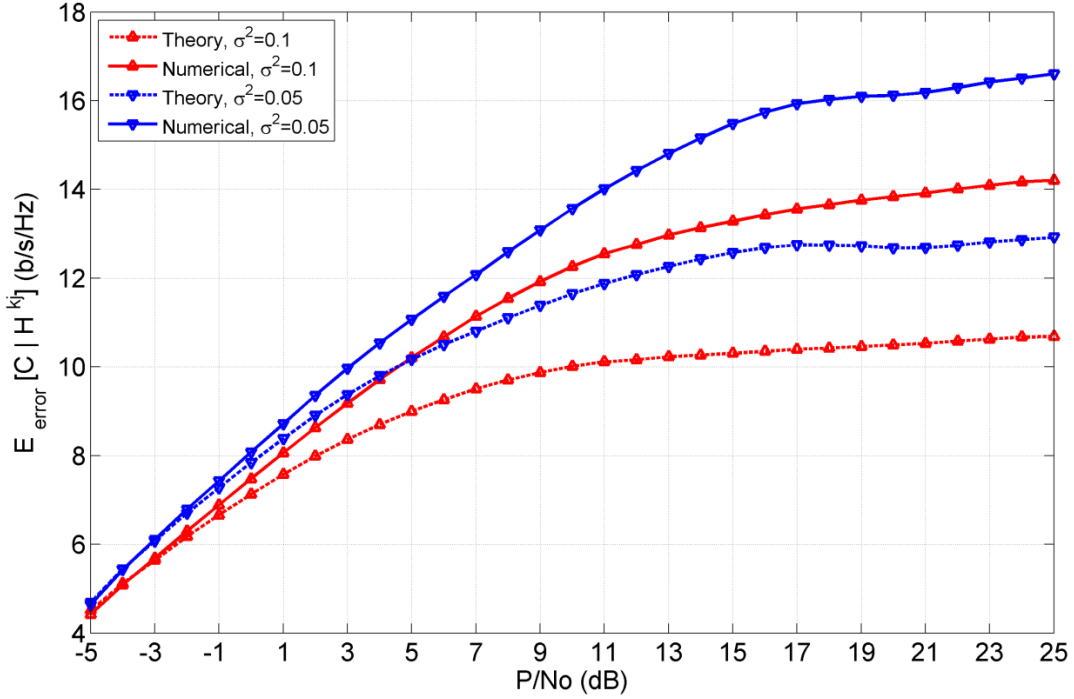

**Fig. 8.** *Theoretical prediction of $E[C|H^{kj}]$ and numerical value for $(3 \times 3,1)^4$ MIMO IC is shown versus SNR. The filters are designed with EM algorithm and two CSI error variances $\sigma_1^2 = 0.05$ and $\sigma_2^2 = 0.1$.*

### 5.5. Convergence of VM Algorithm

Since it is hard to prove convergence of this problem theoretically (It is mathematically intractable) it is investigated numerically. Convergence of the iterative algorithm is shown numerically by considering fraction of interference leakage to the received signal parameter [22]. Fig. 9 shows parameter for the proposed schemes versus iterations. In $(3 \times 3,1)^4$ MIMO IC, EM and VM converge after 10 and 50 iterations as Fig. 9 shows. Proposed algorithms converge after 20 iterations in $(4 \times 4,2)^3$ MIMO IC.

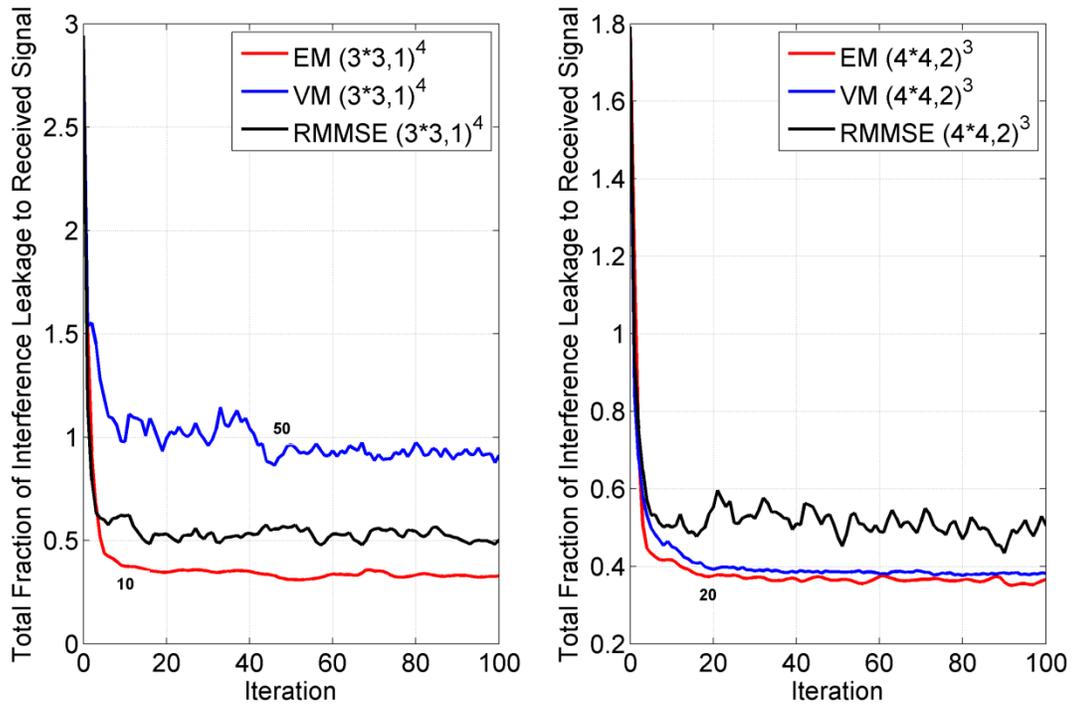

***Fig. 9.*** *Sum of fraction of interference leakage to received signal for two interference networks with $\sigma^2 = 0.1$, and $\frac{P}{N_0} = 10dB$. Convergence behavior of VM is like other algorithms in two considered MIMO IC scenarios. Parameter decreases and then remains within a neighborhood of final value.*

The stopping criterion for the convergence of the iterative algorithms is 100 iterations. Further results regarding the convergence of the VM in larger MIMO ICs are provided for $(10 \times 10,5)^3$ and $(6 \times 8,4)^2$ in Fig. 10. Proposed algorithms converge after 10 iterations (left hand figure). EM and VM converge after 10 and 30 iterations in $(6 \times 8,4)^2$ MIMO IC.

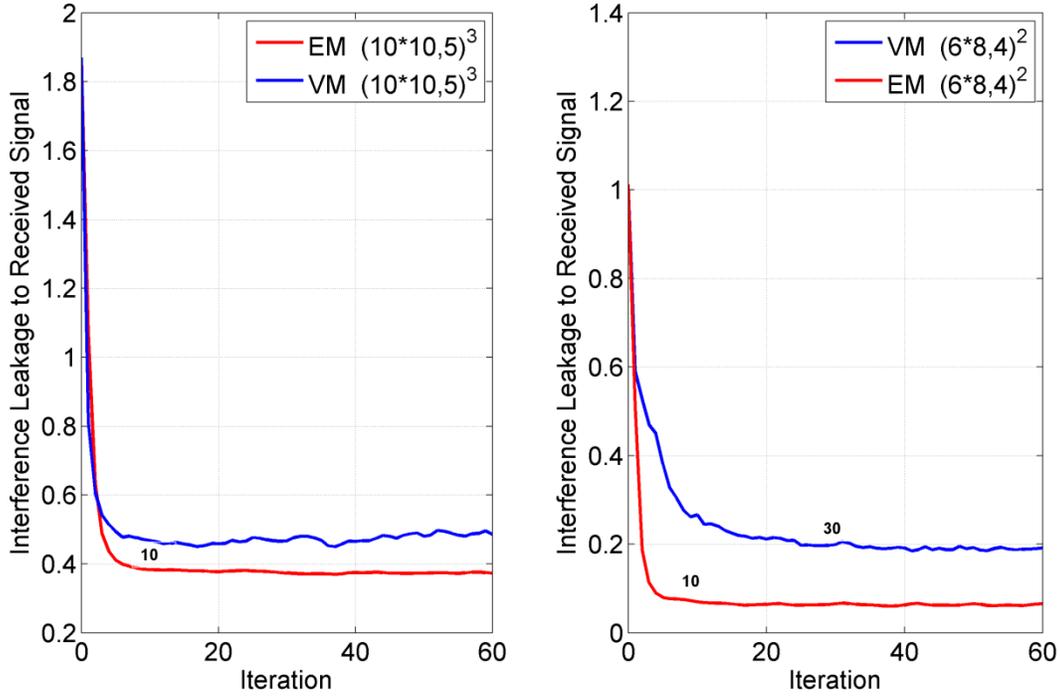

*Fig. 10. Sum of fraction of interference leakage to received signal in $(10 \times 10,5)^3$ and $(6 \times 8,4)^2$ MIMO ICs with $\sigma^2 = 0.1$, and $\frac{P}{N_0} = 10dB$. Parameter decreases after each iteration and then remains constant around final value.*

## 6. Conclusion

In this paper, two robust algorithms were proposed. In the EM scheme, filters were adjusted based on the problem of expectation maximization of SINR. The other design minimized the variance of SINR to hedge against variability due to the CSI error. Taylor series expansion was exploited to approximate the effect of imperfection in CSI on statistical properties. Proposed robust algorithms utilized the reciprocity of wireless networks to optimize estimated statistical properties in two different working modes. Monte Carlo simulations demonstrated that the EM scheme improves data rate of MIMO IC under imperfect CSI. The VM algorithm provided SINR with low variance. Moreover, it improved sum rate, but not as satisfactory as the EM scheme.

## Appendix A

### Estimate the Mean of $SINR_d^k$

If $f(num, den)$ is concentrated near its mean, then $E\left[SINR_d^k = \frac{num}{den}\right]$ can be expressed in terms of $\mu_1$ and $\mu_2$, as the mean values of $num$ and $den$, respectively. $\mu_1$ represents the conditional expected value of $num$

$$\mu_1 = E[num|H^{kj}] = u_d^{k\dagger}[T_d^k + P\sigma^2 I]u_d^k, \tag{31}$$

and $\mu_2$ denotes the conditional expected value of $den$

$$\mu_2 = E[den|H^{kj}] = u_d^{k\dagger}[S^k - T_d^k + (P\sigma^2 \sum_{j=1}^K D^j - P\sigma^2 + N_0)I]u_d^k.^7 \tag{32}$$

According to the statistical linearization argument [29], $SINR_d^k$ is approximated by a first order Taylor series expansion around mean value $(\mu_1, \mu_2)$:

$$SINR_d^k(num, den) \cong SINR_d^k(\mu_1, \mu_2) + \frac{\partial SINR_d^k(\mu_1,\mu_2)}{\partial num}(num - \mu_1) + \frac{\partial SINR_d^k(\mu_1,\mu_2)}{\partial den}(den - \mu_2) \tag{33}$$

.

In this case, (6) yields

$$E[SINR_d^k] \cong \frac{\mu_1}{\mu_2} + \frac{\partial SINR_d^k(\mu_1,\mu_2)}{\partial num}\int\int(num - \mu_1)f(num, den)d.num \times d.den \tag{34}$$

---

[7] In (31) and (32), the following equality has been used $E\left[E^{kj}v_m^j v_m^{j\dagger} E^{kj\dagger}\right] = \sigma^2\left(v_m^{j\dagger}v_m^j\right)I = \sigma^2 I.$

$$+\frac{\partial SINR_d^k(\mu_1,\mu_2)}{\partial den}\int\int(den-\mu_2)f(num,den)d.num \times d.den \ .$$

The value of integrations in (34) is zero. Therefore, estimation of the mean value can be expressed by $E[SINR_d^k] \cong \frac{\mu_1}{\mu_2}$.

# Appendix B

# Max-SINR: Special Case of Proposed EM Algorithm

This appendix starts with a relation which is needed to proof Max-SINR is a special case of proposed EM algorithm. Searle identity [30] for matrix $B$ and column vector $b$ is

$$(B-bb')^{-1}b = \frac{B^{-1}b}{1-b'B^{-1}b} \ . \tag{35}$$

This identity is applied to the column of receive interference suppression matrix of the Max-SINR algorithm (relation 22 in [13]) and it is simplified as follow

$$u_d^k = \frac{(B^k-bb^\dagger+N_0 I)^{-1}G^{kk}v_d^k}{\|(B^k-bb^\dagger+N_0 I)^{-1}G^{kk}v_d^k\|} = \frac{(B^k+N_0 I)^{-1}G^{kk}v_d^k}{\|(B^k+N_0 I)^{-1}G^{kk}v_d^k\|}, \tag{36}$$

where $bb'$ is covariance matrix of $d^{th}$ desired data stream at receiver $k$ and $B^k$ is the covariance matrix of all data streams heard. Matrices are with respect to perfect CSI.

$$B^k = P\sum_{j=1}^{K}\sum_{m=1}^{D^j}G^{kj}v_m^j v_m^{j\dagger}G^{kj\dagger} \ ,$$

$$bb^\dagger = PG^{kk}v_d^k v_d^{k\dagger}G^{kk\dagger} \ . \tag{37}$$

Since for $\sigma^2 = 0$ we have $\Omega_d^k = N_0$ and $G^{kj} = H^{kj}$, it can be concluded that receive filters of EM design are receive matrices of Max-SINR.

# Authors Declarations

*Competing interests*

The authors declare that they have no competing interests.

*Funding*

There are not any sources of funding for the research to be declared.

*Authors' contributions*

AD carried out the preliminary studies, participated in the proposed algorithm, and drafted the manuscript. Dr. MK carried out the figure design and adjustment parameters. DR. HA conceived of the study and participated in its design and coordination and helped in drafting the manuscript. All authors read and approved the final manuscript.

*Acknowledgements*

There is not anyone who contributed towards the article to be acknowledged by authors.